\documentclass[11pt]{article}
\usepackage{eacl2017}
\usepackage{times}
\usepackage{url}
\usepackage{latexsym}
\usepackage{amsmath}
\usepackage{graphicx}
\usepackage{tabu}
\usepackage{float}
\usepackage{booktabs}
\eaclfinalcopy

\newcommand{\at}[2][]{#1|_{#2}}

\title{Exploration of Proximity Heuristics in Length Normalization}

\author{Pranav A \\
	VIT University \\
	Vellore, India \\
	{\tt cs.pranav.a@gmail.com} }

\date{}

\begin{document}
	\maketitle
	\begin{abstract}
		Ranking functions used in information retrieval are primarily used in the search engines and they are often adopted for various language processing applications.
		However, features used in the construction of ranking functions should be analyzed before applying it on a data set. 
		This paper gives guidelines on construction of generalized ranking functions with application-dependent features. 
		The paper prescribes a specific case of a generalized function for recommendation system using feature engineering guidelines on the given data set. 
		The behavior of both generalized and specific functions are studied and implemented on the unstructured textual data. 
		The proximity feature based ranking function has outperformed by 52 \% from regular BM25.
	\end{abstract}
	
	\section{Introduction}
	
	Information retrieval deals with the task of \textit{retrieving}, or in simple words, obtaining relevant and necessary information resources from a huge collection of documents. The information obtained is considered relevant according to a piece of information asked about (also known as \textit{query}) \cite{zhai2016text}. The use of information retrieval has been fundamental for developing search engines. But there are many other interesting applications such as recommendation systems, spam filtering, plagiarism detection and so on \cite{Manning:2008:IIR:1394399}. Ranked information retrieval use ranking functions which determine the decreasing order of the documents in relevance to the query.
	
	\newcite{zhai2016text} lay down the in-depth analysis of variety of ranking functions used in information retrieval. Common features used in the information retrieval are term frequency, inverse document frequency and length normalization. TF-IDF (Term Frequency and Inverse Document Frequency) is a prevalent vector-space information retrieval technique. It balances the similarity of the query and the document, and penalizes the common terms \cite{tfidf}. Pivoted document length normalization with TF-IDF helps to reward shorter documents \cite{pl}. Okapi BM25 is a complex version of pivoted length normalization and it is the prevailing state-of-the-art retrieval method \cite{bm25}. PL2 is a retrieval model based on divergence from randomness of the query term frequency \cite{PL2}. Dirichlet Prior based retrieval is based on language modeling  where the smoothing function is derived from Dirichlet distribution \cite{dirich}. MPtf2ln and MDtf2ln are improvements on previous methods and more elaborate ranking functions which balances the extremities of normalization effects \cite{Fang}. \newcite{Fang} also lay down certain guidelines to evaluate the behavior of the ranking functions.
	
	This paper analyses the feature engineering concepts in the information retrieval, especially length normalization. The objective of this paper are:
	
	\begin{itemize}
		\item to give general guidelines of general structure and nature of feature which can be included in the ranking function. (Section 2)
		\item to study the existing work which use information retrieval in recommendation systems. (Section 3)
		\item to prepare an unstructured textual dataset for penpal recommendation system. (Section 4)
		\item to analyze the dataset and apply the general guidelines to construct the feature for penpal recommendation system. (Section 5)
		\item to test and compare the constructed function against other ranking functions. (Section 6)
	\end{itemize}
	
	\section{Guidelines for Feature Engineering in Ranking Functions}
	
	The main idea of the information retrieval systems is to include a proximity score in conjunction to the ranking function. A proximity score is an application-dependent score, like page-rank metrics, contextual similarity and so on. The function should have well engineered feature embedded in itself and it should be able to generalize other results. The objectives of the ranking function are:
	
	\begin{itemize}
		\item The score should be higher if the constituents of proximity score (real numbers or vectors) are similar.
		\item The proximity score should be below certain cut-off scores. These cut-off scores or limiting parameters can be determined through machine learning.
		\item The function should behave similar and able to generalize the behavior and working of the existing ranking functions in IR.
	\end{itemize}
	
	The assumption is taken that modeled function is bi-modal (as one has to consider extremities of the cut-off scores). Let $f(\mathbf{x}, \mathbf{y})$ be the modeled function, where $\mathbf{x}$ and $\mathbf{y}$ be the constituent vectors. In other words, this function determines similarity between $\mathbf{x}$ and $\mathbf{y}$ and it is a function of $d(\mathbf{x})$ and $d(\mathbf{y})$ where $d(\cdot)$ is the distance metric.

	Length normalization functions, especially in the BM25 or pivoted length function are inversely proportional to the to the TF-IDF score. Thus, the model would be $f(\mathbf{x}, \mathbf{y})$ inversely proportional to the TF-IDF score.

	\begin{equation*}
		TF-IDF(q, d) \propto \frac{1}{f(\mathbf{x}, \mathbf{y})}
	\end{equation*}
	
	\subsection{Nature of the Function Curve}
	
	\begin{figure}[h]
		\centering
		\includegraphics[width=0.43\textwidth]{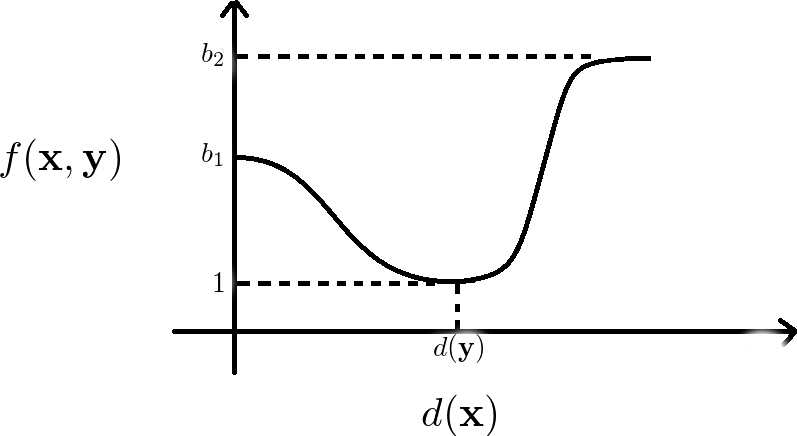}
		\caption{Rough Sketch of Feature Curve}
		\label{sketch}
	\end{figure}

	Since the function has to be bi-modal distribution, distribution function would be modeled in the form asymmetrical inverted bell curve. Thus the trough of the curve should occur when distances of $\mathbf{x}$ and $\mathbf{y}$ are similar. In other words,
	
	\begin{equation} \label{c1}
		\frac{\partial f(\mathbf{x}, \mathbf{y})}{\partial ( d(\mathbf{x}) )} = 0 \text{ if } d(\mathbf{x}) \approx d(\mathbf{y})
	\end{equation}
	
	Similarly it can be shown that,
	
	\begin{equation*}
		\frac{\partial f(\mathbf{x}, \mathbf{y})}{\partial (d(\mathbf{y}))} = 0 \text{ if } d(\mathbf{x}) \approx d(\mathbf{y})
	\end{equation*}
	
	From here onwards, arguments would be taken from the $d(\mathbf{x})$ perspective, as $d(\mathbf{y})$ would have similar arguments.
	
	The function should decrease monotonically if $d(\mathbf{x}) < d(\mathbf{y})$. In other words, 
	
	\begin{equation} \label{c2}
		\frac{\partial f(\mathbf{x}, \mathbf{y})}{\partial ( d(\mathbf{x}) )} < 0 \text{ if } d(\mathbf{x}) < d(\mathbf{y})
	\end{equation}
	
	The function should increase monotonically if $d(\mathbf{x}) > d(\mathbf{y})$. In other words,
	
	\begin{equation} \label{c3}
		\frac{\partial f(\mathbf{x}, \mathbf{y})}{\partial ( d(\mathbf{x}) )} > 0 \text{ if } d(\mathbf{x}) > d(\mathbf{y})
	\end{equation}
	
	\subsection{Nature of the Limits of the Curve}
	
	Here the left extremity of the curve could be bounded to the parameter $b_1$.
	\begin{equation} \label{c4}
		\lim_{d(\mathbf{x})\to 0} f(\mathbf{x}, \mathbf{y}) = b_1
	\end{equation}
	
	Similarly, the right extremity of the curve could be bounded to the parameter $b_2$.
	\begin{equation} \label{c5}
		\lim_{d(\mathbf{x})\to \infty} f(\mathbf{x}, \mathbf{y}) = b_2
	\end{equation}
	
	The trough of the curve exists when $\frac{\partial f(\mathbf{x}, \mathbf{y})}{\partial ( d(\mathbf{x}) )} = 0$. Thus the limit is set to 1. It can not be set it to 0, because $f(\mathbf{x}, \mathbf{y})$ lies in the denominator of the scoring function.
	\begin{equation} \label{c6}
		\lim_{d(\mathbf{x})\to d(\mathbf{y})} f(\mathbf{x}, \mathbf{y}) = 1
	\end{equation}
	
	The specific case of these guidelines can be applied to the recommendation systems.
	
	\section{Information Retrieval in Recommendation Systems}
	
	Not many efforts have been carried out which use ranking functions in recommendation systems. Most of the recommendation systems use latent Dirichlet allocation (LDA) or collaborative filtering.  \newcite{nextgen} give a more comprehensive study of existing research work on recommendation systems. 
	
	Some recommendation systems have experimented with ranking functions. However, all of them did not tune the parameters correctly. The systems which deploy IR functions, have generally used extremely well-structured data like tags \cite{course} or contextual variables \cite{Kwon}. 
	
	\textbf{Tag-based systems :} These systems have concatenated the tags of an user into a single \textit{document}. Relevant proximity is determined through similarity between \textit{query} and tags. This idea has used to develop music-recommendation systems \cite{Cantador,Bellogin}. The cosine similarity between BM25 value of user profile tags and item profile tags produced the best results. This leads to conclusion that BM25 outperforms other functions because BM25 penalizes common tags and focuses on rare terms \cite{Cantador}. The effects of length normalization are not discussed. More evaluation metrics have been introduced to evaluate the ranking functions \cite{Bellogin}. Results concluded that collaborative filtering methods have more novel and diverse recommendations and ranking functions have more coverage and better accuracy. TF-IDF outperforms BM25 but the parameters of BM25 have not been tuned. 
	
	\textbf{Publication-recommendation systems :} Some publication-recommendation systems have used ranking functions \cite{twitter}. Concept Frequency Inverse Document
	Frequency (CF-IDF) and  Hierarchical CF-IDF (HCF-IDF) have been utilized which generally assign weights on certain pre-determined words in the documents. The results showed that CF-IDF combined with sliding window produced the best results. Surprisingly, it outperforms the popular LDA method. Another publication-recommender system have used information retrieval elements \cite{Totti}. Here citation context (TF-IDF similarity between two papers), query similarity (TF-IDF similarity between query and article) and age decay (to penalize older articles) have been considered as parameters. This experiment also showed that it surpasses the system that uses page-rank like metrics.
	
	\textbf{Recommendation systems that use unstructured data:} \newcite{Suchal} and \newcite{Esparza} have used unstructured textual data in the recommendation system designs. Users' choices have been concatenated into a single query which is used to recommend articles \cite{Suchal}. Movie recommendation has also been evaluated with ranking functions \cite{Esparza}. Tags and reviews of movies are served as an TF-IDF component. Reviews tend to have better performance than tags because unstructured nature  of reviews provide some noise and undiscovered information as opposed to structured data like tags. This results high IDF values. Thus unstructured textual data is a suitable advantage. According to the results, BM25 underperformed against TF-IDF algorithm. This happened because parameters have not been tuned according to the dataset. \newcite{Esparza} also demonstrated that ranking functions perform better than collaborative filtering algorithms.
	
	The main conclusions from this literature survey are:
	
	\begin{enumerate}
		\item Ranking functions tend to perform better than well-known recommendation system algorithms like LDA and collaborative filtering.
		\item Unstructured data provides certain amount of noise which is helpful for ranking function parameters.
		\item Effects of length normalization on recommendation system have not been studied.
	\end{enumerate}
	
	\section{Penpal Recommendation System and Dataset Preparation}
	
	Penpal (online friends) recommendation system was set up from the online users. 630 users were asked about their interests, likes and relationships. Few volunteers helped in assessing in the matching of the penpals. Since every respondent was exclusively assigned one penpal, thus it gave 315 pairs as a result. 
	
	The response of the user was concatenated into a single sample or \textit{document}. Hence the recommendation system situation was there for textual data for 630 users. MeTA \cite{massung2016meta} toolkit was employed in the codebase of this project. Preprocessing on the data was done with the stopword removal, tokenization and lemmatization.  The dataset was divided into two parts:
	
	\begin{itemize}
		\item \textbf{Training set} : Text data of 504 samples (252 pairs)
		\item \textbf{Testing set} : Text data of 126 samples (63 pairs)
	\end{itemize}
	
	Average document length of the training set is 131 words. The observations followed that assessors paired up users whose response contained fewer words and similar interests (Table \ref{table}). Similar process was carried out for users having lengthy responses and similar interests. This may be due to the reason that user who is writing lesser words in the response form is less interested in having a penpal. Thus text similarity and length similarity plays a major role in penpal recommendation.
	
	\begin{table}[h]
		\begin{tabu} to 0.5\textwidth { | X[l] | X[c] | }
			\hline \bf Description & \bf Number of Pairs \\
			\hline 
			Total Pairs in Training Dataset & 252 \\ \hline
			Pairs matched with both document's length less than the average document length  & 62\\ \hline
			Pairs matched with both document's length than the average document length  & 131\\ \hline
			Others matches & 53\\ \hline
		\end{tabu}
		\caption{\label{table} Document length characteristics in the training dataset }
	\end{table}
	
	\section{Feature Engineering}
	Here, the function would be modeled using Richard's curve. The curve is defined by,
	\begin{equation}
		g(x) = l + \frac{u - l}{(A + e^{{-B(x-M)}})^{{1/\nu }}}
	\end{equation}
	
	Here, $l$ is lower limit, $u$ is upper limit and $M, \nu, A, B$ are free parameters
	
	The distance metrics are taken to be just real numbers. Thus, $d(\mathbf{x}) = x$ and $d(\mathbf{y}) = y$. The trough of the curve should occur when, $d(\mathbf{x}) = d(\mathbf{y})$ or $x = y$. The function can be partitioned into three parts:
	\begin{itemize}
		\item monotonically decreasing Richard's curve, $g(x)$  when $x < y$
		\item monotonically increasing Richard's curve, $g(x)$ when $x > y$
		\item Value of heuristic function, $h(x, y) = 1$  when $x = y$
	\end{itemize}
	
	The definition of the function now looks like:
	
	\begin{equation}
		h(x, y) = 
		\begin{cases}
			g(x_1)       & \quad \text{if } x < y\\
			1 & \quad \text{if } x = y\\
			g(x_2)  & \quad \text{if } x > y\\
		\end{cases}
	\end{equation}
	
	where $x_1 \in [0, y]$, $x_2 \in [y, \infty]$ and $x_1, x_2 \subset x$
	
	Applying equation (\ref{c1}), the constraint obtained is:
	
	\begin{equation*}
		\frac{\partial h(x, y)}{\partial x}\at[\bigg]{x=y} = 0
	\end{equation*}
	
	From equation (\ref{c2}), the constraint obtained is:
	
	\begin{equation*}
		\frac{\partial g(x_1)}{\partial x_1} < 0
	\end{equation*}
	
	From equation (\ref{c3}), the constraint obtained is:
	
	\begin{equation*}
		\frac{\partial g(x_2)}{\partial x_2} > 0
	\end{equation*}

	From limiting condition (\ref{c4}), the constraint obtained is: 
	
	\begin{align*}
		\lim_{x_1 \to 0} g(x_1) &= b_1 \\
		\frac{\partial g(x_1)}{\partial x_1}\at[\bigg]{x_1 = 0} &\approx 0
	\end{align*}
	
	From the limiting condition (\ref{c5}), the constraint obtained is: 
	
	\begin{align*}
		\lim_{x_2 \to \infty} g(x_2) &= b_2 \\
		\frac{\partial g(x_2)}{\partial x_2}\at[\bigg]{x_2 = \infty} &\approx 0
	\end{align*}
	
	From the limiting condition (\ref{c6}), the constraint obtained are: 
	
	\begin{align*}
		\frac{\partial g(x_1)}{\partial x_1}\at[\bigg]{x_1 = y} &\approx \frac{\partial g(x_2)}{\partial x_2}\at[\bigg]{x_2 = y} &\approx 0  \\
		\lim_{x_1 \to y} g(x_1) &= \lim_{x_2 \to y} g(x_2) &= 1 
	\end{align*}
	
	After solving the constraints, and using the binomial approximation for small $\nu$, the solutions obtained are:
	
	\begin{equation}
		h(x, y) = 
		\begin{cases}
			1 + \frac{b_1 - 1}{1 + e^{B_1(x - cy)}}       & \quad \text{if } x < y\\
			1 & \quad \text{if } x = y\\
			1 + \frac{b_2 - 1}{1 + e^{-B_2(x - (1 + c)y)}}  & \quad \text{if } x > y\\
		\end{cases}
	\end{equation}
	
	where,
	
	\begin{itemize}
		\item $B_1$ and $B_2$ are growth parameters, can be typically set to 1
		\item $c$ is the trough curvature, where $c \in (0, 1)$
	\end{itemize}
	
	According to the requirements, length similarity would be rewarded between query and document lengths. In that case, $x = |d|$ and $y = |q|$. Intuitively, that means the value of the scoring function will be higher if $|d| \approx |y|$.
	
	Substituting the values, we get:
	
	\begin{equation}
		h(|d|, |q|) = 
		\begin{cases}
			1 + \frac{b_1 - 1}{1 + e^{B_1(|d| - c|q|)}}       & \text{if } |d| < |q|\\
			1 &  \text{if } |d| = |q|\\
			1 + \frac{b_2 - 1}{1 + e^{-B_2(|d| - (1 + c)|q|)}}  &  \text{if } |d| > |q|\\
		\end{cases}
	\end{equation}
	
\begin{figure}[h]
		\centering
		\includegraphics[width=0.48\textwidth]{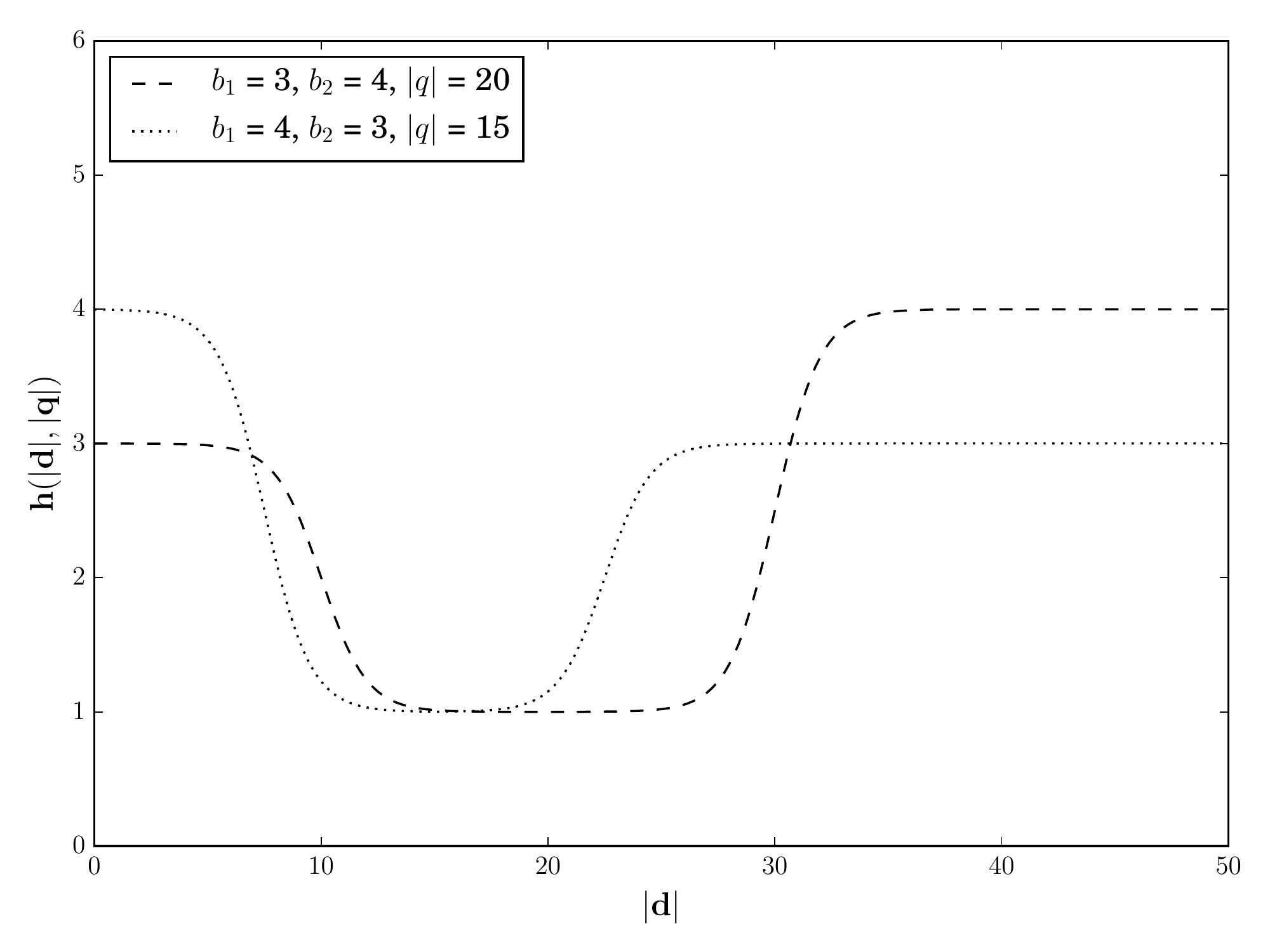}
		\caption{The nature of $h(|d|, |q|)$ with $B_1 = 1$, $B_2 = 1$ and $c = 0.5$}
		\label{nature}
	\end{figure}

	The nature of this feature can be plotted as shown against various parameters (Figure \ref{nature}). It can be seen that it closely resembles the desired graph.
	
	This feature can now be substituted in the ranking function in place of length normalization function. For example, the original BM25 function is:
	
	\begin{align*}
		score(q, d) = & \sum_{t \in d \cap q} f(t, q) \log \frac{M + 1}{df(t)} \times \\
		& \frac{ (k + 1) f (t, d) }{f(t, d) + k \left( 1 - b + b \frac{|d|}{avgdl} \right)} 
	\end{align*}
	
	The normalization feature $\left( 1 - b + b \frac{|d|}{avgdl} \right)$ in the BM25 formula can be replaced with $h(|d|, |q|)$ to get our desired ranking function:
	\begin{equation}
		\label{heuris}
		\begin{aligned}
			score(q, d) = & \sum_{t \in d \cap q} f(t, q) \log \frac{M + 1}{df(t)} \times \\
			& \frac{ (k + 1) f (t, d) }{f(t, d) + k \left( h(|d|, |q|) \right)}  \\
		\end{aligned}
	\end{equation}
	\section{Results and Conclusions}
	The relevant feedback of the dataset is limited. The human assessors have provided only one relevant document for  one query. Thus, only MRR (Mean Reciprocal Rank) is used to assess the models.
	
	\begin{equation}
		MRR = \frac{1}{s} \sum_{i = 1}^s \frac{1}{rank_i}
	\end{equation}
	
	Here $s$ is the size of the dataset and $rank_i$ is the position of the rank for the first relevant document of the $i^{th}$ query.
	
	The parameters of the baseline algorithms were tuned according to the training dataset and tested on the test set using MRR. 
	
	\begin{table}[H]
		\begin{center}
			\begin{tabular}{ | m{2.56cm} | m{1.5cm}| m{1.5cm} | } 
				\hline
				\textbf{Model} & \textbf{Training Set MRR} & \textbf{Testing Set MRR} \\
				\hline
				Length Similarity Heuristic with BM25  & 0.34  & 0.29  \\
				\hline
				BM25  & 0.24  & 0.19  \\
				\hline
				Pivoted Length Normalization & 0.22  & 0.18  \\
				\hline
				MPtf2ln & 0.21  & 0.18  \\
				\hline
				MDtf2ln  & 0.19  & 0.16  \\
				\hline
				PL2 & 0.17  & 0.15  \\
				\hline
				Dirichlet Prior & 0.16  & 0.13  \\
				\hline
			\end{tabular}
			\caption{Mean reciprocal rank (MRR) values on the data-set using various retrieval methods}
			\label{results}
		\end{center}
	\end{table}
	
	The obtained MRR values have been given in (Table \ref{results}). The proximity feature has outperformed by 52 \% from regular BM25. It can be observed that query-document length similarity, not document length normalization has helped in this situation. 
	
	\begin{table}[H]
		\centering
		\begin{tabular}{|c|c|}
			\hline
			\multicolumn{1}{|c|}{\textbf{Parameter}} & \multicolumn{1}{c|}{\textbf{Value}} \\ \hline
			k (BM25 parameter)                                        & 2.8                                 \\ \hline
			$b_1$ (Left Bound)                                    & 2.9                                 \\ \hline
			$b_2$ (Right Bound)                                   & 3.7                                 \\ \hline
			$B_1$ (Growth parameter)                                   & 1                                   \\ \hline
			$B_2$  (Growth parameter)                                  & 1                                   \\ \hline
			c (Trough curvature)
			& 0.5                                 \\ \hline
		\end{tabular}
		\caption{Parameters used in similarity feature in BM25}
		\label{my-label}
	\end{table}
	
	The parameters used in similarity feature (Table \ref{my-label}) with BM25 show that $b_1$ and $b_2$  are tuned around 3 and 4. This means that the magnitude of TF-IDF value is penalize around 3 or 4 times when the lengths are dissimilar.
	
	Further work is to be carried out by using more evaluation tests, creating a better dataset, giving proofs for generalization and applying the algorithm into more applications like spelling correction.

	\bibliography{eacl2017}
	\bibliographystyle{eacl2017}

\end{document}